\documentclass[final]{svjour2}
\usepackage{graphicx}
\usepackage{rotating}
\usepackage{amssymb, amsmath}
\usepackage{mathptmx}
\usepackage[numbers]{natbib}
\usepackage{color}

\usepackage[nodisplayskipstretch]{setspace}
\makeatletter
\journalname{Journal of Low Temperature Physics}
\bibpunct{}{}{,}{s}{}{,}

\begin{document}

%\listoffigures

\newcommand{\hdblarrow}{H\makebox[0.9ex][l]{$\downdownarrows$}-}
\newcommand{\mnu}{m(\nu_\text{e})}

\title{Algorithms for Identification of Nearly-Coincident Events in
Calorimetric Sensors}

\author{B.~Alpert \and E.~Ferri \and D.~Bennett \and
 M.~Faverzani \and J.~Fowler \and A.~Giachero \and J.~Hays-Wehle \and M.~Maino
\and A.~Nucciotti \and A.~Puiu \and D.~Swetz \and J.~Ullom}

\institute{Contribution of U.S. government not subject to copyright in
the United States\\
\\
B.~Alpert \and D.~Bennett \and J.~Fowler \and J.~Hays-Wehle
 \and D.~Swetz \and J.~Ullom\\
National Institute of Standards and Technology, Boulder, Colorado 80305, USA\\
\email{alpert@boulder.nist.gov}\\
\\
E.~Ferri \and M.~Faverzani \and A.~Giachero \and M.~Maino
 \and A.~Nucciotti \and A.~Puiu\\
Universit\`a degli Studi di Milano-Bicocca \and INFN Sez.~di Milano-Bicocca,
Milan, Italy}

\date{\today}

\maketitle

\vspace{-0.13in}
\begin{abstract} For experiments with high arrival rates, reliable
  identification of nearly-coin\-ci\-dent events can be crucial.  For
  calorimetric measurements to directly measure the neutrino mass such
  as HOLMES, unidentified pulse pile-ups are expected to be a leading
  source of experimental error.  Although Wiener filtering can be used
  to recognize pile-up, it suffers errors due to pulse-shape variation
  from detector nonlinearity, readout dependence on sub-sample arrival
  times, and stability issues from the ill-posed deconvolution problem
  of recovering Dirac delta-functions from smooth data.  Due to these
  factors, we have developed a processing method that exploits
  singular value decomposition to (1) separate single-pulse records
  from piled-up records in training data and (2) construct a model of
  single-pulse records that accounts for varying pulse shape with
  amplitude, arrival time, and baseline level, suitable for detecting
  nearly-coincident events.  We show that the resulting processing
  advances can reduce the required performance specifications of the
  detectors and readout system or, equivalently, enable larger sensor
  arrays and better constraints on the neutrino mass.

  \vspace{-0.05in}

\end{abstract}

\begin{figure}[t]
  \includegraphics[scale=0.585, bb=30 0 588 320]{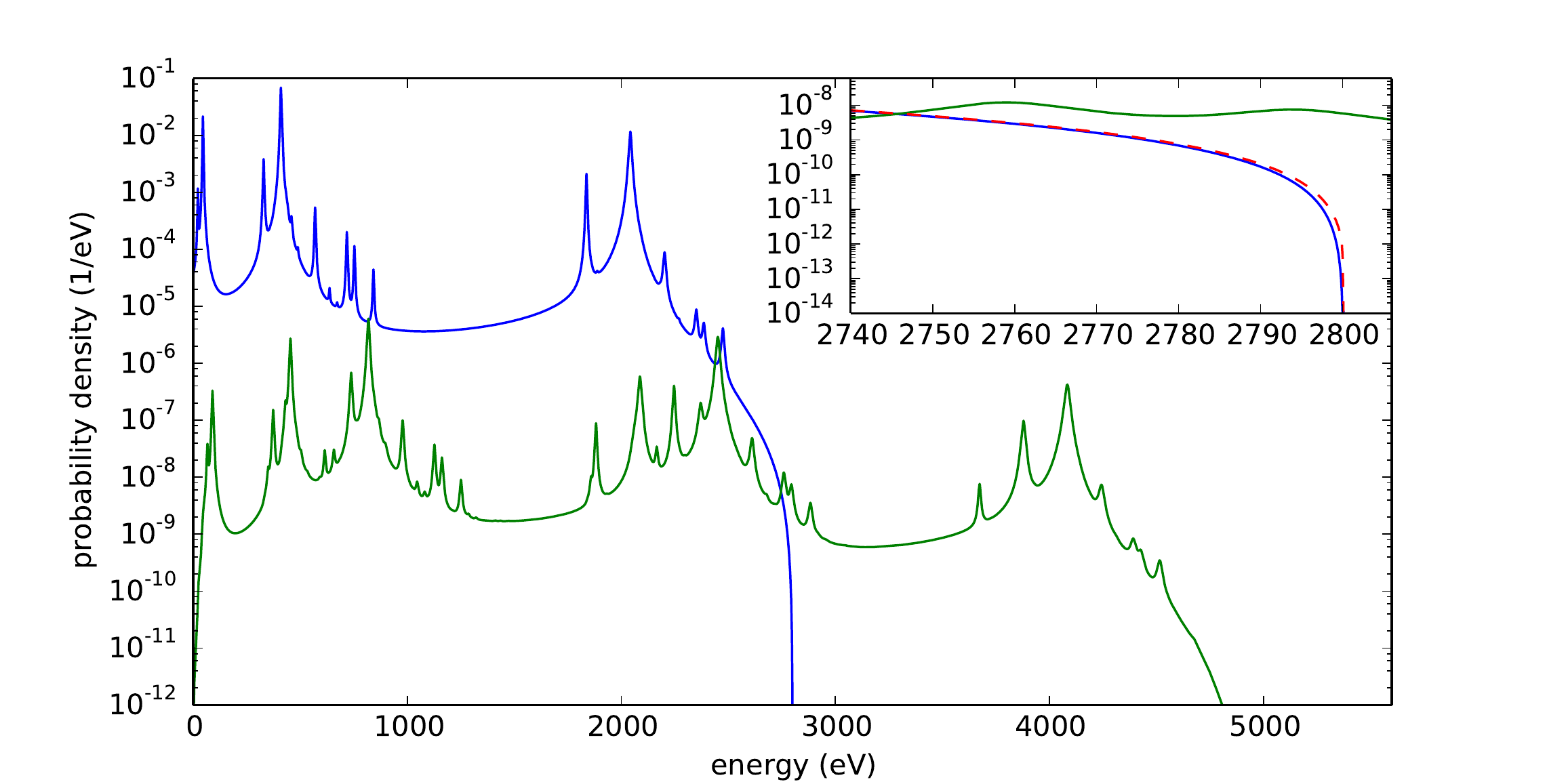}
  \caption[fig1]{De-excitation spectrum from $^{163}$Ho electron
    capture, based on one- and two-hole states [4], energy endpoint
    $Q=2800$ eV, and neutrino mass $\mnu=0$ eV, is shown (blue) along
    with its self convolution (green), the spectrum of a piled-up pair.
    Latter is scaled by the relative probability $3.00\times 10^{-4}$
    of pile-up from an event rate of 300 /s and time resolution of 1
    $\mu$s.  Inset focuses on energies of interest near $Q$ and adds
    single-event spectrum for $Q=2801$, $\mnu=1$ eV (red, dashed).  In
    an experiment, just the sum of single-pulse and pile-up spectra is
    observable. (Color figure online.)}
  \label{spectra}
\end{figure}

\section{Introduction}
Experiments attempting direct measurement of the electron
anti-neutrino mass $\mnu$, including ECHo [1], HOLMES [2], and NuMECS
[3], which measure the de-excitation energy of $^{163}$Dy produced by
$^{163}$Ho electron capture, seek to quantify the spectrum of this
de-excitation energy, after neutrino escape, near its endpoint
$Q-\mnu$ where its shape is sensitive to $\mnu$.  HOLMES experiment
specifications, for example, are for 300 events per detector per
second with time resolution of 1 $\mu$s to separate distinct events,
but under these conditions misleading piled-up events with energy sum
in the vicinity of $Q-\mnu$ are more prevalent than single events with
similar energy; see Fig.~\ref{spectra}.  Minimizing erroneous
counting of double events as single is crucial to the success of these
experiments.

A TES microcalorimeter, heated by the energy of an arriving particle,
traverses a sharp transition in resistance from its bias point near
superconductivity toward normal resistance; this change is observed as
a drop in current through the device, with pulse onset and decay rates
determined largely by circuit inductance and thermal conductance to a
heat bath.  A rapid pulse rise, facilitating pile-up detection, is one
strategy to reduce undetected pile-up of pulses, but at a given sample
rate there is a trade-off between pulse rise time and effective energy
resolution.  Another strategy is optimal (Wiener) filtering [5] to
invert the effect of detector linear impulse response and additive
colored Gaussian noise.  Obstacles in practice, however, of detector
nonlinear response, pulse rising-edge readout distortion tied to event
sub-sample arrival time [6], and inherent instability of Wiener filter
construction due to uncertainty of pulse shape and noise power
spectrum, motivate exploration of alternative methods.

Simulations demonstrate the efficacy of a new approach based on
singular value decomposition (SVD), both to distinguish piled-up pulse
records from more numerous single-pulse records in training data and
to construct a model of the single-pulse records for use in
identifying relatively much more numerous piled-up records near the
energy-spectrum endpoint.  For sample spacing $\Delta t$ we observe
time resolution near $\Delta t/2,$ which we argue below is ideal.  By
comparison, the time resolution of Wiener filtering is observed [7]
as approximately $\Delta t$.

\section{Energy Spectrum, Detector Model, and Processing Procedure}
\vspace{-0.06in}
\subsection{Energy spectrum}
Electron capture by $^{163}$Ho yields excited $^{163}$Dy whose
de-excit\-ation energy spectrum (Fig.~\ref{spectra}) has probability
density
\begin{equation}
  \frac{\text{d}\Gamma}{\text{d}E}=(Q-E)\sqrt{(Q-E)^2-{\mnu}^2}\sum_f
\frac{\lambda_0B_f\Gamma_f}{2\pi\left[(E-E_f)^2+{\Gamma_f}^2/4\right]}
\label{density}
\end{equation}
well established theoretically [8]. Total energy $Q$ and triples
$E_f,B_f,\Gamma_f,$ specifying center, amplitude, and width of the
Lorentzians comprising terms of the summation, have been determined
experimentally and computationally, with recent major improvements
in determining line locations and widths [9],
second- and third-order transitions [4,10,11,12], and $Q$ [13].

In the following we omit third-order electron transitions from simulation.

\vspace{-0.1in}
\subsection{Detector model}
For this study we simulate current pulses in a transition-edge-sensor
(TES) microcalorimeter.  The dynamics of detector temperature $T$ and
current $I$ are modeled by ordinary differential equations
\begin{align}
  C\frac{\text{d}T}{\text{d}t}&=-k\cdot (T^n-{T_\text{bath}}^n)+I^2R(T,I)+
                  \sum_i\delta\left(t-t_i\right)\cdot E_i\label{eqT}\\
  L\frac{\text{d}I}{\text{d}t}&=V-I\cdot R_\text{L}-I\cdot R(T,I)\label{eqI}
\end{align}
of the Irwin-Hilton model [14], for energies $E_1,E_2,\ldots,$
arriving at times $t_1,t_2,\ldots,$ with detector resistance given by the
formula
\begin{equation}
  R(T,I)=\frac{R_\text{N}}{2}\left[1+\text{tanh}\left(\frac{T-T_\text{c}+(I/A)^{2/3}}{2\text{ln}(2)T_\text{w}}\right)\right]
\label{ShankR}
\end{equation}
proposed by Shank et al. [15], with the small-signal linearizations
[14] of the transition---logarithmic temperature sensitivity $\alpha$
and logarithmic current sensitivity $\beta$ of detector resistance $R$
at quiescence---replaced by the functional form of Eq.~(\ref{ShankR}).
Physical parameters in Eqs.~(\ref{eqT})--(\ref{ShankR}) are chosen to be
similar to those of detectors being developed [16] at NIST for HOLMES,
specifically\\
\vspace{-0.20in}

\begin{center}
\begin{tabular}{lll}
\multicolumn{3}{l}{Assumed:}\\
$n=3.25$ & $T_\text{c}=0.1$ K & $T_\text{bath}=0.07$ K\\
$k=23.3$ nW/K$^n$ & $C=0.5$ pJ/K & $L\in\{12,24,48\}$ nH\\
$R_0=2$ m$\Omega$ & $R_\text{L}=0.3$ m$\Omega$ & $R_\text{N}=10$ m$\Omega$\\
$\alpha=\partial\ln R/\partial\ln T\big|_0=200.0$ & $\beta=\partial\ln R/\partial\ln I\big|_0=2.0$ & \\
\multicolumn{3}{l}{Derived:}\\
$T_0=0.0980$ K & $I_0=63.85\;\mu$A & $G=406.8$ pW/K \\
$T_\text{w}=0.565$ mK & $A=1.133$ A/K$^{3/2}$ & $V=146.9$ nV.\\
\end{tabular}
\end{center}

\vspace{-0.05in}
\noindent
Temperature at quiescence $T_0$ is obtained from $\alpha,$ $\beta,$ $R_0,$ 
and Eq.~(\ref{ShankR}).
Current at quiescence $I_0$ is then obtained from temperature balance
and parameters $T_\text{w}$ and $A$ are obtained from $T_0,$ $I_0,$
$R_0,$ $\alpha,$ and $\beta$, while
$G=\partial\left[ k\cdot(T^n-{T_\text{bath}}^n)\right]/\partial
T\big|_0$ $=kn{T_0}^{n-1}.$

\begin{figure}[t]
  \includegraphics[scale=0.31, bb=30 0 588 410]{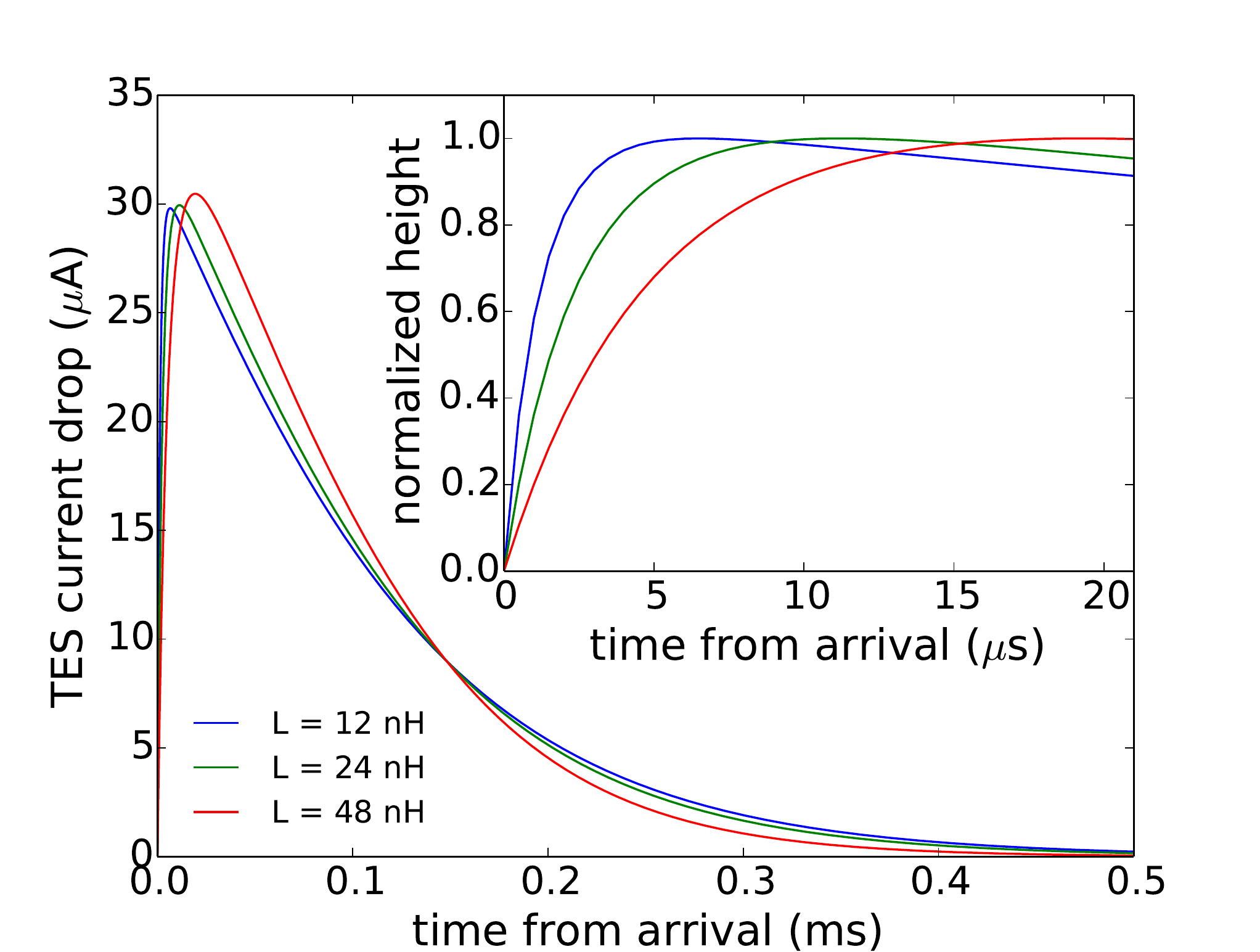}%
  \includegraphics[scale=0.31, bb=30 0 588 410]{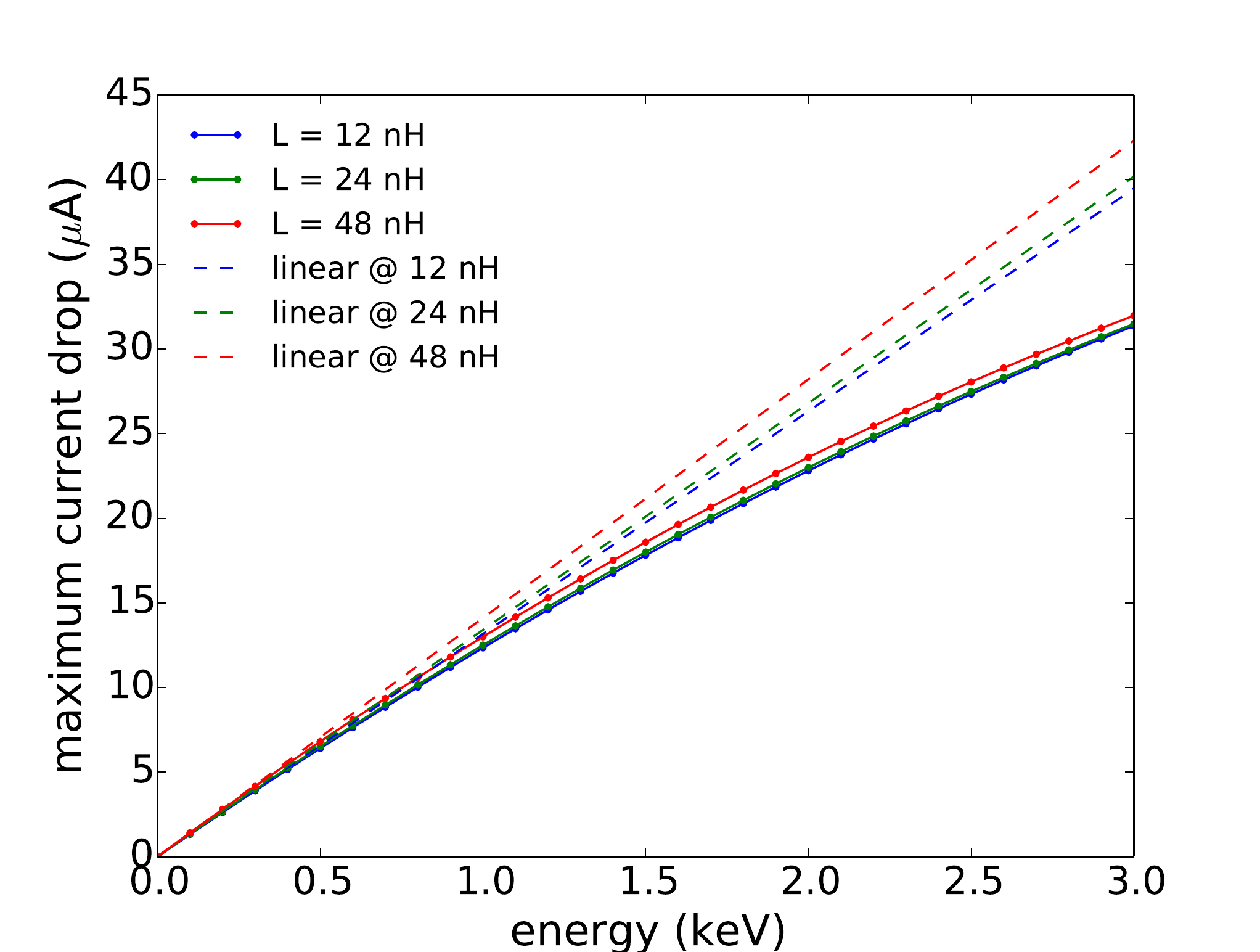}%
  \caption{Pulse shape at 2.8 keV, with rising-edge detail
    ({\it Left}) and linearity of pulse height versus energy with
    slope from 0.1 keV ({\it Right}) for inductances 12, 24, 48 nH.
    The proposed pile-up detection approach achieves near-ideal
    performance on all three.  (Color figure online.)}
  \label{shape-fig}
\end{figure}

Three values of inductance, $L=12,\;24,\;48$ nH, are simulated to
enable evaluation of pile-up detection contrasting relatively rapid
with relatively slow pulse rises, the latter allowing lower sample
rates.  Pulse shape and detector nonlinearity are shown in
Fig.~\ref{shape-fig}.

\begin{figure}[b]
  \includegraphics[scale=0.39, bb=75 0 588 325]{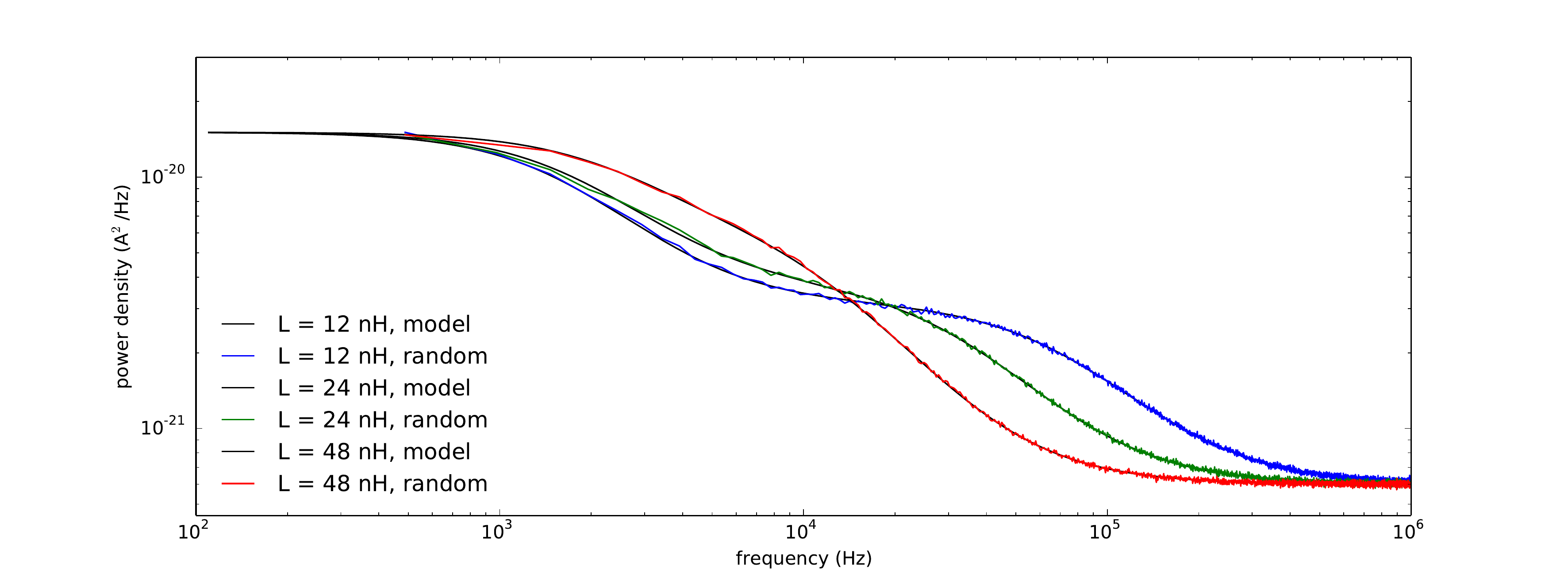}
  \caption{Noise power spectrum---from Irwin-Hilton model and from
    simulation by ARMA process---for inductances $L=12,\;24,\;48$
    nH. (Color figure online.)}
  \label{noise}
\end{figure}

Noise is additive with power spectrum given by the Irwin-Hilton model
[14], which depends on TES circuit inductance, and is generated in the
simulation as an autoregressive moving average (ARMA) process on
Gaussian white noise.  The noise spectra are shown in
Fig.~\ref{noise}.

\vspace{-0.1in}
\subsection{Generation of events and pile-up}\label{prpp}
\vspace{-0.03in} We simulate events according to the one- and two-hole
electron excitation spectrum [4] with endpoint $Q=2.8$ keV (motivated
by the recent independent measurement of $Q$ [13]), neutrino mass
$\mnu=0$~eV, and event mean arrival rate $\lambda=300$ /s. Detecting
pile-up is of particular importance for identifying the single pulses
for energies near $Q,$ which carry the information on $\mnu$.  To
obtain a sufficient number of counts, we restrict generation of each
single-pulse energy $E$ to the interval $S_\text{E}=[2.70,2.82]$ keV.
Piled-up pairs, with lag $t_\Delta$ between arrival times of less than
$\delta=10$ $\mu$s and each event drawn from the full energy spectrum,
subject to the sum of their energies $E_1+E_2$ lying in $S_\text{E},$
have rate fraction
\begin{equation}
f_\text{pp}=\frac{c_\text{pp}}{c_\text{s}+c_\text{pp}}
\label{fppdef}
\end{equation}
where $c_\text{s}$ is the count of single and $c_\text{pp}$ the count of piled-up
pulse pairs.  These incidence frequencies are determined by the
probabilities
\begin{gather*}
\text{Pr}(E\in S_\text{E})=6.78\times 10^{-7},\;\;\text{Pr}(t_\Delta\ge\delta)=
0.997,\\
\text{Pr}(E_1+E_2\in S_\text{E})=2.14\times 10^{-3},\;\;\text{Pr}(t_\Delta<\delta)=
3.00\times 10^{-3},
\end{gather*}
arising from the $^{163}$Ho energy spectrum of Eq.~(\ref{density}) and
the exponential distribution $F(t)=1-\lambda\text{e}^{-\lambda t}$ of
time lags for Poisson arrivals.  Therefore
\begin{equation}
f_\text{pp}=\frac{\text{Pr}(E_1+E_2\in S_\text{E})\text{Pr}(
t_\Delta<\delta)}{\text{Pr}(E_1+E_2\in S_\text{E})\text{Pr}(t_\Delta<\delta)
+\text{Pr}(E\in S_\text{E})\text{Pr}(t_\Delta\ge\delta)}=0.905
\label{fppval}
\end{equation}
and only after detection and rejection are pile-ups less numerous
than single-pulse records.

Once arrival times and event energies are generated, a stream of
samples of current $I$ is obtained by solution of ordinary
differential equations of Eqs.~(\ref{eqT}), (\ref{eqI}) by explicit
fourth-order Runge-Kutta integration on nodes with uniform spacing
$\Delta t=0.5\;\mu$s in addition to nodes at event arrival
times---where model temperature $T$ is discon\-tinuous---and the
uniform nodes are retained.  These samples are subsequently decimated
to obtain sample rates of 2, 1, 0.67, and 0.5 MHz.

\vspace{-0.1in}
\subsection{Recordization}
After ARMA-generated noise is added to the samples, the stream of
samples is formed into pulse records, simulating conditions of the
physical experiment.  Pulse arrivals are determined by a six-sample
pulse trigger in which a line segment is fit to the first five samples
and a threshold is applied to the difference between the sixth sample
and the advanced line height.  This procedure enables detection of a
pulse arrival just five samples after a prior arrival, including on
the rising edge of the prior pulse.  A pulse record of duration 0.5 ms
is formed, provided that 0.1 ms before and 0.4 ms after a pulse
arrival are free of other triggered arrivals.  Records that have
relatively flat pre-trigger are retained and transformed to whiten
noise by a fast Cholesky-factor backsolve procedure [17].

\vspace{-0.15in}
\subsection{Pile-Up detection}
\begin{figure}[t]
  \includegraphics[scale=0.49, bb=55 0 385 345]{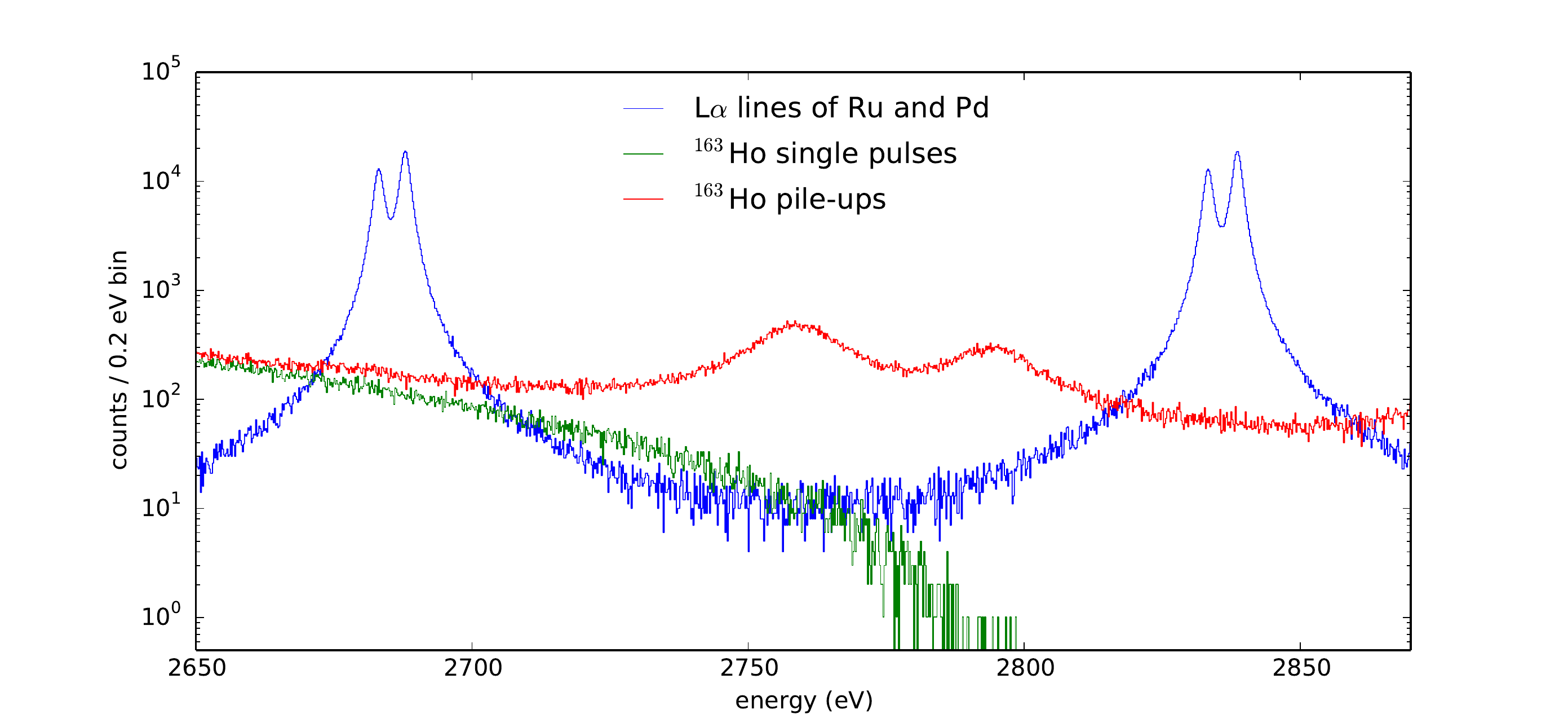}%
  \caption{De-excitation spectra from $^{163}$Ho electron capture have
    single pulse counts (green) heavily dominated by pile-up counts
    (red) near $Q$ in expected physical data and in simulation
    according to Eq.~(\ref{fppval}) (shown).  (Unlike
    Fig.~\ref{spectra}, which assumes 1~$\mu$s time resolution, the
    training data are shown prior to pile-up detection.)  Therefore to
    enable construction of a single-pulse model, in a model training
    phase the $^{163}$Ho source is augmented with a source yielding
    single pulse counts (blue) from L$\alpha$ x-rays of Ru and
    Pd. (Color figure online.)}
  \label{train}
\end{figure}

Our strategy for pile-up detection is to develop a model of
single-pulse records and reject as piled-up those records whose model
fit has too-large residual. The energy interval $S_\text{E}$ has
$^{163}$Ho single-pulse record counts dominated by piled-up-record
counts, as is evident from Eq.~(\ref{fppval}); our models built from
single-pulse-dominated intervals at lower energies (for example, near
the 2.042 keV spectral peak) were not effective on $S_\text{E}$.
Instead we propose adding a switchable source of photons~[18], from
L$\alpha$ x-ray emission lines of Ru (2.683, 2.688 keV) and Pd (2.833,
2.839 keV), likely also of value as reference lines for combining data
of multiple detectors.  A second simulation, for the energy interval
$S'_\text{E}=[2.65,2.87]$ keV, with the switchable source in
combination with $^{163}$Ho to provide approximately five times as
many separated pulses as piled-up pairs, is used as a training set for
developing a single-pulse model (see Fig.~\ref{train}).
\begin{figure}[b]
  \includegraphics[scale=0.52, bb=40 20 385 520]{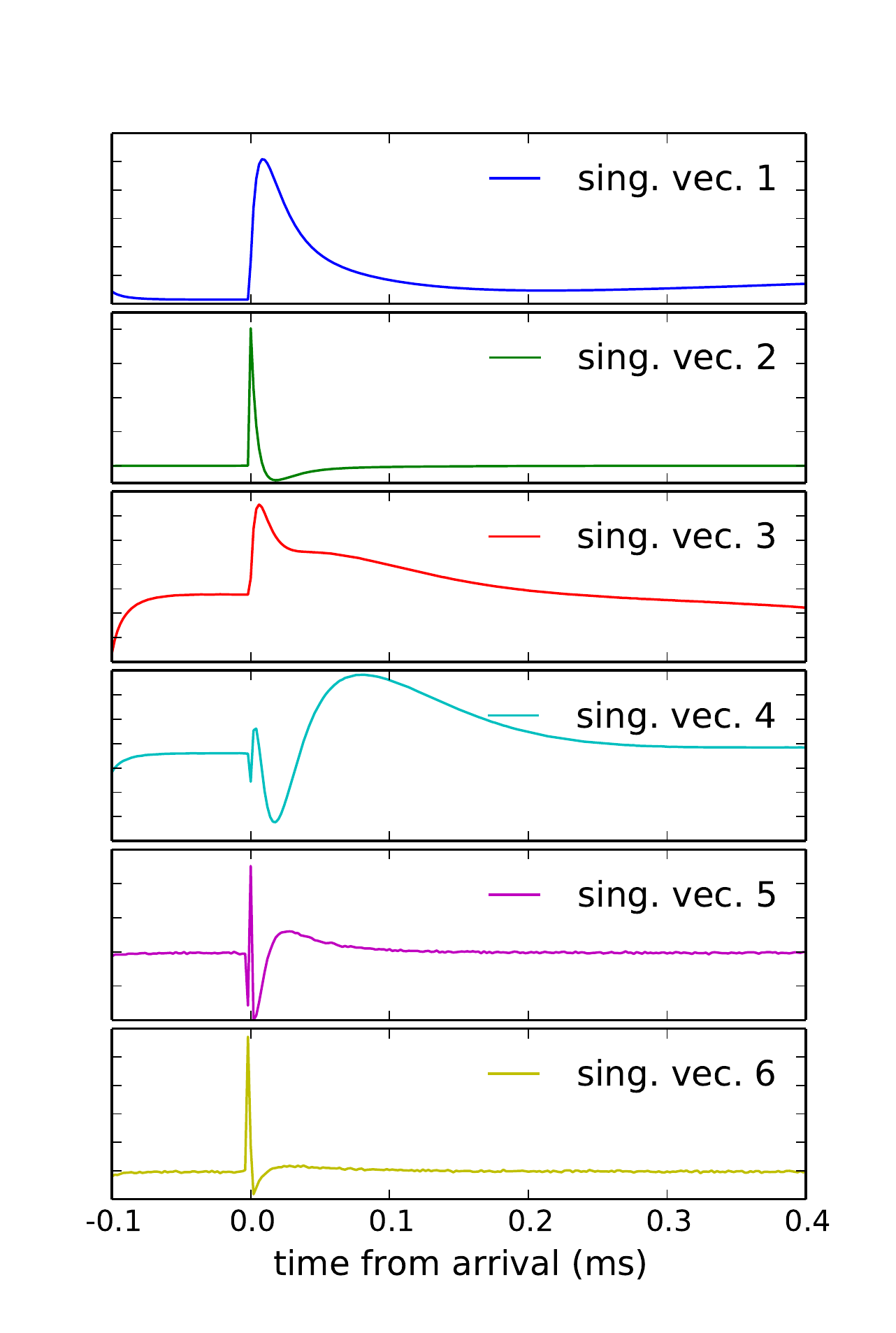}%
  \includegraphics[scale=0.52, bb=40 20 385 520]{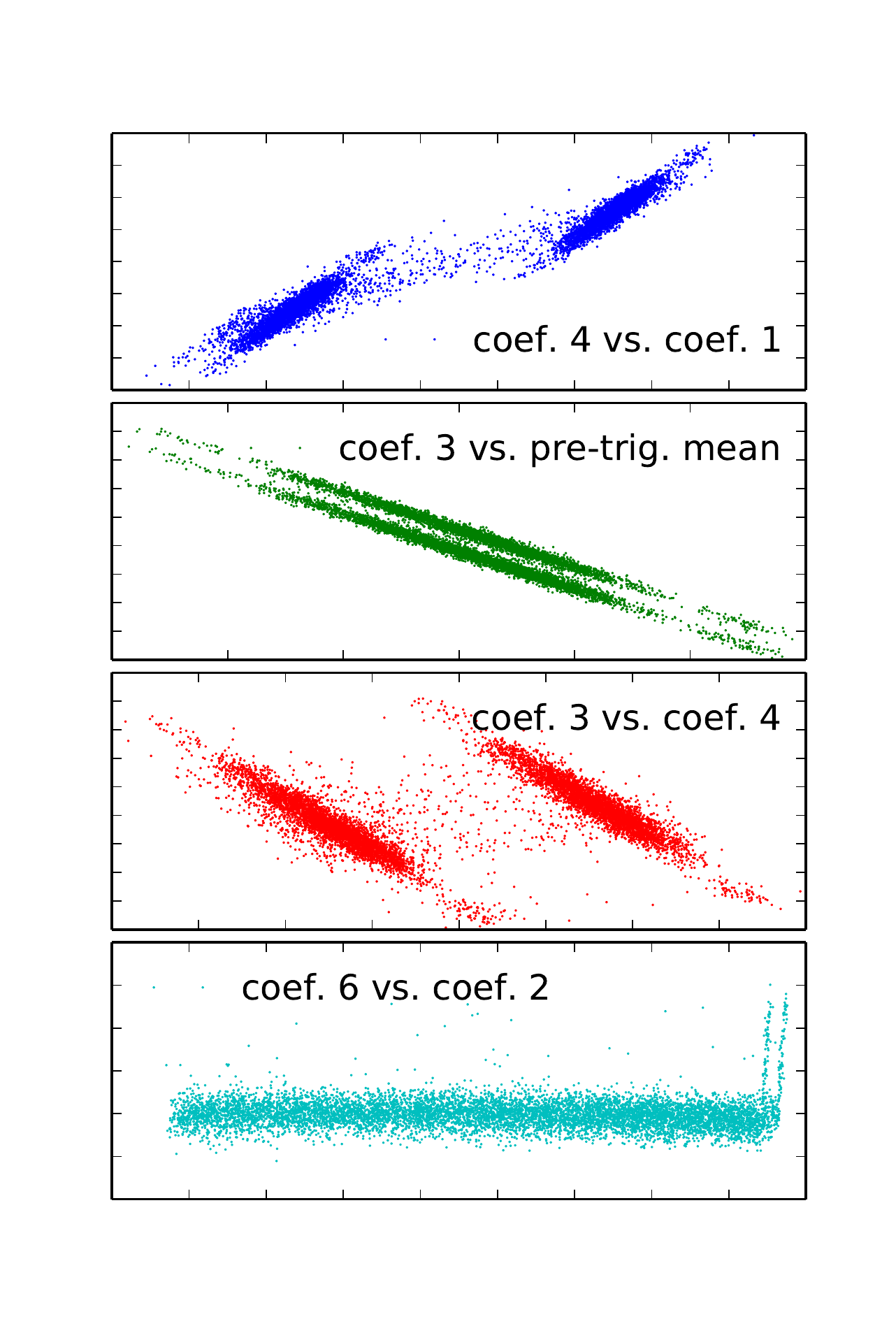}%
  \caption{{\it Left:} Singular vectors linked to largest singular
    values enable representation of variations due to pulse amplitude,
    sub-sample arrival time, and baseline shifts, with nonlinear
    effects, while suppressing noise.  {\it Right:} Certain of their
    coefficients correlate, providing redundancy to allow evaluation
    of pulse record conformance to a single-pulse model.  (Color
    figure online.)}
  \label{singfuncs}
\end{figure}

\subsubsection{Separating single-pulse records}
The model training simulation yields records that, while containing
more single pulses than pile-ups, are not governed by single pulses
adequately to enable direct construction of a single-pulse model.
We therefore precede model construction with a step in which
piled-up records are identified as outliers and removed.

The singular value decomposition (SVD) $M=UDV^t$ is computed for a
matrix $M$ whose columns consist of pulse records
(Fig.~\ref{singfuncs}).  The first $j$ columns ($j<10$) of $U$
comprise a good basis for the full set of records; the remaining
columns are dominated by noise, as signaled by the very slow decay of
the singular values beyond the $j$th.  The means of the first $j$
columns of $V$ are subtracted from those columns to obtain $\hat{V}.$
The $j\times j$ empirical covariance $\hat{\sigma}^2=\hat{V}^t\hat{V}$
is then used to obtain a squared deviation
$d^2=\hat{V}_{i,*}(\hat{\sigma}^2)^{-1}\hat{V}_{i,*}^t$ of each record
$i$.  Piled-up records deviate disproportionately from the mean in
this covariance-adjusted sense so we discard those with largest $d^2,$
and repeat the procedure on the remaining records (form $M,$ compute
SVD $M=UDV^t$ and each $d^2$, discard records with largest $d^2$) a
total of three times, with $m/2$ discards on the first iteration,
$m/4$ on the second, and $m/8$ on the third, where $m$ is the expected
number of piled-up records.  The value of $m$ can be estimated as the
pile-up count in the stream data (as in Eq.~(\ref{fppval}) of
\textsection\ref{prpp}) minus the number of removals by the
recordization process.

The iterations yield bases computed by SVD that successively improve
the representation of single-pulse records, as piled-up records are
increasingly eliminated.  The three iterations and the sequence of
record counts removed per iteration were chosen empirically to
generate a balance between false positives---single-pulse records
discarded as piled up---and false negatives---piled-up records
retained as single pulses.

\begin{table}[t]
  \caption{Initial (pp$_\text{i}$) and final (pp$_\text{f}$) pile-up
    record count fractions of training and evaluation data sets,
    energy resolution $\Delta E$ (FWHM), and effective time resolution
    $\tau_\text{R}$.  Columns labelled F$+$ and F$-$ are the false
    positives---single-pulse records discarded as piled up---and false
    negatives---piled-up records retained as single pulses.  Energy is
    computed by optimal filtering [19] and $\tau_\text{R}$ is the
    ratio of number of retained piled-up records to single-pulse
    records, divided by that ratio of original events ($1\,083\,229$ /
    $114\,049$), times $\delta=10\;\mu$s.}
  \label{dat-table}
\begin{center}
{\small
\begin{tabular}{rrrrrrrrrrr}
\multicolumn{1}{l}{\makebox[0.45in][l]{MHz}} & \multicolumn{4}{c}{count fractions (training)}
 & \multicolumn{1}{c}{$\Delta E$}
 & \multicolumn{4}{c}{count fractions (evaluation)}
 & \multicolumn{1}{c}{$\tau_\text{R}$}\\
%&\multicolumn{1}{c}{$\tau_\text{WF}$}\\
nH & \multicolumn{1}{c}{pp$_\text{i}$}
 & \multicolumn{1}{c}{F$+$} & \multicolumn{1}{c}{F$-$}
 & \multicolumn{1}{c}{pp$_\text{f}$} & (eV)
& \multicolumn{1}{c}{pp$_\text{i}$}
 & \multicolumn{1}{c}{F$+$} & \multicolumn{1}{c}{F$-$}
 & \multicolumn{1}{c}{pp$_\text{f}$} & ($\mu$s)\\ % & ($\mu$s)\\
\multicolumn{1}{l}{2.0}\\
12 & .042 & .002 & .172 & .007 & 1.99 & .681 & .009 & .139 & .231 & 0.29 \\
24 & .048 & .002 & .167 & .008 & 2.10 & .707 & .013 & .225 & .355 & 0.31 \\
48 & .054 & .002 & .162 & .009 & 2.31 & .724 & .011 & .243 & .391 & 0.49 \\
\multicolumn{1}{l}{1.0}\\
12 & .073 & .002 & .153 & .012 & 2.46 & .796 & .010 & .140 & .355 & 0.55 \\
24 & .081 & .002 & .151 & .013 & 2.45 & .813 & .010 & .131 & .366 & 0.56 \\
48 & .087 & .005 & .173 & .016 & 2.60 & .824 & .011 & .136 & .391 & 0.60 \\
\multicolumn{1}{l}{0.67}\\
12 & .101 & .005 & .174 & .019 & 3.34 & .849 & .008 & .145 & .451 & 0.83 \\
24 & .109 & .003 & .153 & .018 & 2.76 & .859 & .009 & .133 & .450 & 0.81 \\
48 & .112 & .005 & .168 & .021 & 2.80 & .864 & .007 & .142 & .476 & 0.84 \\
\multicolumn{1}{l}{0.5}\\
12 & .127 & .009 & .185 & .027 & 7.54 & .880 & .006 & .150 & .525 & 1.11 \\
24 & .130 & .007 & .175 & .026 & 3.17 & .883 & .007 & .142 & .518 & 1.09 \\
48 & .124 & .009 & .189 & .026 & 2.89 & .877 & .006 & .160 & .534 & 1.08 \\
\end{tabular}
}
\end{center}
\end{table}

\subsubsection{Constructing single-pulse model}
The separation procedure culls the collection of training records so
that nearly all of those remaining are single-pulse records, which
enables construction of a single-pulse model.  The SVD $M=UDV^t$ is
again computed and the first $j$ of the expansion coefficients
$(VD)_{i,*}$ for record $i$ are augmented with the record's
pre-trigger mean.  The first column of $U$ approximates the average
pulse, while the second column is dominated by the effect of varying
arrival time on pulse shape.  The remaining basis vectors encode
variations due to changing baseline and to nonlinear effects of
varying pulse height, arrival time, and baseline, so we approximate
expansion coefficients $3,\ldots,j$ by linear regression from a
nonlinear space from coefficients 1, 2, and the pre-trigger mean.  In
particular, naming these three independent variables $x,\;y,\;z,$ we
construct a model for each of the $j-2$ dependent variables as a
linear combination of $1,\;x,\;y,\;z,\;xy,\;yz,\;zx,\;xyz,$ resulting
in $8\cdot (j-2)$ model coefficients.

The model residual is obtained for each training record and, more
importantly, for any pulse record, when $j$ expansion coefficients
(along with pre-trigger mean) are computed by taking the inner product
of the record with the basis vectors.  Records with residual norm
above a threshold are deemed piled-up, where the threshold is chosen
to include 99~\% of the training set pulse records that survive as
single pulsed.  A lower threshold would yield only a few more detected
pile-ups.

For both training steps, of separating single-pulse records and of
constructing the single-pulse record model, we choose $j=6$.  Readout
distortion on the pulse rising edge [6]---not modeled here---likely
will imply $j>6$ in the physical setting.

\subsubsection{Energy bias of pile-up detection}
After pile-up detection and rejection, many records with pile-up
remain---false negatives---and some single-pulse records are
discarded---false positives.  Correct analysis of the experiment
therefore requires assessing the extent of energy bias inherent in the
pile-up detection procedure.

Initially attempting graphically to estimate the bias, as a function
of energy, of the detection procedure, we find good agreement between
the distributions of accepted and rejected records across the energy
interval $S_\text{E}$.

We then compare the energy distribution of retained single-pulse
records to that of all other generated single pulses, and the
distribution of retained piled-up records---with the two energies
summed---to that of all other generated pile-ups, by means of the
two-sample Kolmogorov-Smirnov distribution-equality test.

\vspace{-0.1in}
\section{Simulation Results}
Table~\ref{dat-table} shows initial and final pile-up record count
fractions, and false positive and false negative count fractions, for
training and evaluation data.  In addition, the
energy resolution $\Delta E$ and effective time resolution
$\tau_\text{R}$ of the procedure are shown.  We note that pulses of
the 12 nH inductance models---with fastest rise---are poorly resolved
at the lower sampling rates.  This leads to poor energy resolution
$\Delta E$, but the time resolution $\tau_\text{R}$ after pile-up
detection suffers only slightly.  Also, not shown here, varying the
noise level affects $\Delta E$ significantly, but $\tau_\text{R}$ just
slightly.

\begin{figure}[t]
  \includegraphics[scale=0.35, bb=85 0 588 420]{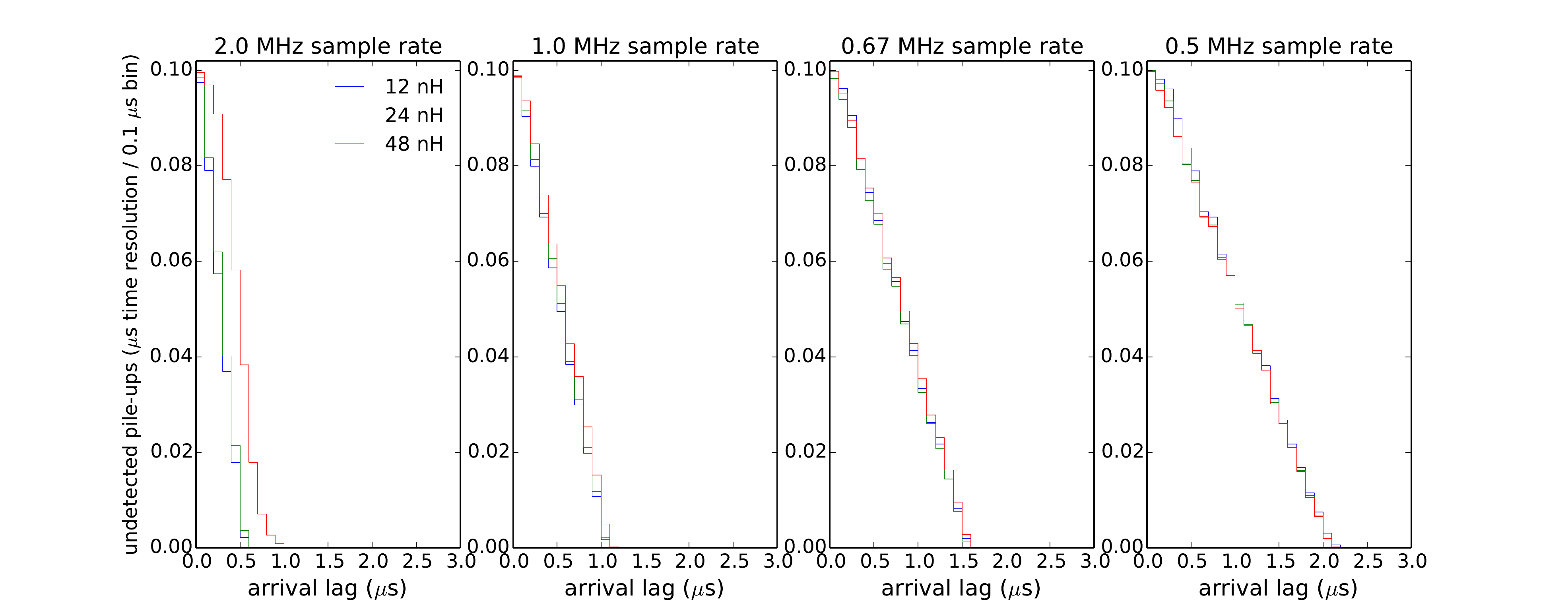}
  \caption{Undetected piled-up record counts are shown as a function
    of time lag $t_\Delta$ between arrivals of the first and second
    events of a pair. (Vertical scale results from 0.1 $\mu$s
    arrival-lag bin size.)  One sample between arrivals generally
    enables their detection as a pair, resulting in the nearly linear
    dependence seen. (Color figure online.)}
  \label{perf-fig}
\end{figure}

Fig.~\ref{perf-fig} shows the undetected piled-up record counts as a
function of event arrival lag $t_\Delta$ for the 12 cases.  These
plots make clear that, regardless of pulse shape variation and noise,
the performance of the pile-up detection depends closely on whether or
not there is at least one sample between two event arrivals.  Achieved
time resolution $\tau_\text{R}$ shown in Table~\ref{dat-table} can be
compared with values of 0.25, 0.50, 0.75, and 1.00~$\mu$s---half the
sample spacing---for sampling rates of 2.0, 1.0, 0.67, and 0.5~MHz,
respectively.  This criterion characterizes ideal performance, since
under the temperature-current model of Eqs.~(\ref{eqT})--(\ref{eqI}), two
arrivals with no intervening sample cannot be distinguished from a
single arrival with energy roughly the sum of the two energies.

The Kolmogorov-Smirnov test (at 95 \% level) of energy bias in
retaining sin\-gle-pulse records failed to reject the null
hypothesis---equivalent to no bias---for all 12 data sets.
The test for energy bias in retaining piled-up records failed to
reject the null hypothesis for 7 of the 12 data sets, which suggests
energy bias on 5 of the 12.  The bias, like the energy resolution and
unlike the time resolution, is sensitive to the noise level, with
lower noise producing less bias.  Preferable to a hypothesis test
would be determination of the effect of bias on the final estimate of
the quantity of interest $\mnu$, which is important for other pile-up
detection methods as well.  We undertake this analysis and present its
results elsewhere.

\vspace{-0.1in}
\section{Conclusions}
These simulations show that there is approximately a two-fold
reduction in pile-up, after rejection, compared with Wiener filtering.
The improvements provided by these algorithms, based on SVD, have
significant, beneficial implications for planned and future
experiments that employ calorimetric sensors to measure the neutrino
mass.  For current experiments with a defined number of pixels such as
HOLMES, this improved pile-up rejection enables a lower sampling rate
per detector, thus relaxing the requirements on the multiplexing
readout system and subsequent detector speeds, while providing similar
or improved reduction in uncertainty due to pile-up and therefore
better constraint on the neutrino mass.  In future experiments, where
the measurement uncertainty due to pile-up will be better understood, a
lower sample rate will allow more detectors to be multiplexed per
readout channel for the specified time resolution requirement.  This
capability reduces the total number of necessary multiplexing channels
and will allow for larger arrays of detectors in a single cryogenic
cooler, thus reducing both the cost and complexity and enhancing the
feasibility of such experiments.

\vspace{0.1in}\noindent {\footnotesize {\bf Acknowledgements}
  The NIST Innovations in Measurement Science program and the European
  Research Council, funding HOLMES, supported this work.  The authors
  also thank Kevin Coakley, Galen O'Neil, and Hideyuki Tatsuno for
  helpful comments on a draft of this paper.}

%\vspace{-0.15in}

\end{document}